\documentstyle[12pt]{article}
\begin{document}
\title{FIRST-ORDER TRANSITIONS IN CHARGED
BOSON NEBULAE} \vspace{1cm}
\author{Ciprian Dariescu
and Marina-Aura Dariescu\thanks{On leave of
absence from Department of Theoretical Physics,
{\it Al. I. Cuza} University, Bd. Copou no. 11,
6600 Ia\c{s}i, Romania,
email (after May 1, 2003): marina@uaic.ro} \\
{\it Institute of Theoretical Science} \\
{\it 5203 University of Oregon, Eugene, OR 97403}\\
email: marina@physics.uoregon.edu}
\date{}

\maketitle

\begin{abstract}
Using the first-order approximating solutions to the
Einstein-Maxwell-Klein-Gordon system of equations for a complex
scalar field minimally coupled to a spherically symmetric
spacetime, we study the feedback of gravity and electric field on
the charged scalar source. Within a perturbative approach, we
compute, in the radiation zone, the transition amplitudes and the
coherent source-field regeneration rate.
\end{abstract}
\vspace*{0.2cm}
{\it Keywords}:
\begin{itemize}
\item{}
Klein-Gordon-Maxwell-Einstein equations;
\item{} charged boson stars.
\end{itemize}

\baselineskip 1.5em
\newpage

\newpage

Boson stars, generally seen as gravitationally bound, both
globally $U(1)$ and spherically symmetric, compact equilibrium
configurations of cold complex scalar fields, were discovered
theoretically, more than 30 years ago, by Kaup [1] and Ruffini and
Bonazzola [2]. Later, a major interest has been focused on
macroscopic stable boson stars, once they arose as promising
candidates for non-baryonic dark matter in the universe [3].
Moreover, according to current observations, it has been suggested
that a supermassive boson or soliton star could be at the center
of our galaxy [4]. These assumptions raise the question whether
such configurations, if they exist, are dynamically stable [5] and
the settling down of a boson star to a stable state has been
investigated via numerical calculations [6].

In our approach, we term by a boson star the charged scalar nebula
which finds itself in one of the spherically symmetric
positive-frequency modes of radial wave number $k$, with $k^2 \to
0_+$, and pulsation $\omega_k = \left[ m_0^2 + k^2 \right]^{1/2}$.
Such a configuration is obviously unstable and the instabilities
is expected to lead to the formation of boson star from a
initially smooth state [7]. Not very much is specifically known in
this direction [8], although the reversed stability to instability
passage has been extensively investigated [5, 6, 9].

In what it concerns an analytical approach, exactly solvable
models for boson stars with large selfinteraction [10] and for
boson-fermion stars [11] have been worked out only in low
dimensional gravity. In four dimensions, the bosonic or the mixed
fermion-bosonic fields interacting via gravity have been
investigated mainly by numerical calculations [1-6,8,9,12].
Therefore, the analytical solutions of the coupled field equations
could be of interest for a better understanding of different
stellar configurations as well as for a numerical-functional
combined iterative treatment which describes the dynamics of
charged boson stars formation.

In a previous letter [7], we have studied the $SO(3,1) \times
U(1)$-gauged minimally coupled charged spinless field to a
spherically symmetric spacetime and analytically obtained the
first-order approximating solutions to the system of
Klein-Gordon-Maxwell-Einstein equations. These allow us to go
further and to study now the feedback of gravity and electric
field on the charged scalar source, computing perturbatively, in
the long range approximation, the correspondingly induced
transition amplitudes.

Let us consider the spherically symmetric configuration described
by the metric
\begin{equation}
ds^2 \, = \, e^{2 f} \, (dr)^2 \, + \, r^2 \,
\left( d \theta^2 + \sin^2 \theta \, d \varphi^2 \right)
- \, e^{2 h} \, (dt)^2 \, ,
\end{equation}
where $f$ and $h$ are functions of $r$ and $t$,
and define the pseudo-orthonormal tetradic frame
$\lbrace e_a \rbrace_{a=\overline{1,4}}$
with the corresponding dual orthonormal base
\begin{equation}
\omega^1 \, = \, e^f dr \; ,
\; \; \omega^2 \, = \, r \, d \theta \; ,
\; \; \omega^3 \, = \, r \sin \theta \, d \varphi \; ,
\; \; \omega^4 \, = \, e^h \, d t
\end{equation}

The $SO(3,1) \times U(1)$-gauge invariant Lagrangean density
for a charged boson of mass $m_0$,
coupled to the electromagnetic field,
\begin{equation}
{\cal L} \, = \, \eta^{ab} \, \bar{\phi}_{;a} \phi_{;b}
\, + \, m_0^2 \, \bar{\phi} \phi \, +
\, \frac{1}{4} \, F^{ab} F_{ab} \, ,
\end{equation}
lead to the following system of
Klein-Gordon-Maxwell-Einstein equations
\begin{eqnarray}
& &
\Box \phi \, - \, m_0^2 \, \phi \, = \,
2 ie \, A^c \phi_{|c} \, + \, e^2 A^c A_c \phi
\; , \; \; \; \; \;
{\rm and \; its \; h.c. } \, , \\*
& &
{F^{ab}}_{:b} \, = \, - \, ie \, \eta^{ab}
\left[ \bar{\phi} \left( \phi_{|b}
- ie A_b \phi \right) - \left( \bar{\phi}_{|b} + ie A_b
\bar{\phi} \right) \phi \right]
\\*
& &
G_{ab} \, = \, \kappa \left(
\bar{\phi}_{;a} \phi_{;b} \, + \,
\bar{\phi}_{;b} \phi_{;a} \, + \,
F_{ac} F_b^{\; \; c} \, - \, \eta_{ab} {\cal L} \right)
\end{eqnarray}
The components of the Einstein tensor, $G_{ab}$,
have the explicit form
\begin{eqnarray}
G_{11} & = & 2 \, \frac{e^{-f}}{r} \, h_{|1} \, - \,
\frac{1-e^{-2f}}{r^2}
\nonumber \\*
G_{22} & = & G_{33} \, = \nonumber \\*
& = &
\, h_{|11} \, + \,
\left( h_{|1} \right)^2
\, - \, \left[ f_{|44} + \left( f_{|4} \right)^2
\right] + \, \frac{e^{-f}}{r} \, h_{|1} \, + \,
\left[ \left( \frac{e^{-f}}{r} \right)_{|1} +
\frac{e^{-2f}}{r^2} \right]
\nonumber \\*
G_{44} & = & - \, 2 \, \left[ \left(
\frac{e^{-f}}{r} \right)_{|1} \, + \,
\frac{e^{-2f}}{r^2} \right] + \, \frac{1-e^{-2f}}{r^2}
\nonumber \\*
G_{14} & = & 2 \, \frac{e^{-f}}{r} \, f_{|4}
\end{eqnarray}
where $( \, \cdot \, )_{|a} = e_a ( \, \cdot \, )$.
Also, in (3),
$\phi_{;a} \, = \, \phi_{|a} - ie A_a \phi$
and the Maxwell tensor $F_{ab} = A_{b:a} - A_{a:b}$
is expressed in terms of the
Levi-Civita covariant derivatives
of the four-potential $A_a$, i.e.
$A_{a:b} = A_{a|b} - A_c \Gamma^c_{ab}$.

In our previous letter [7], we have assumed that the charged
scalar field is the main source of both the electromagnetic and
gravitational fields and neglected (at large distances and within
the framework of a first-order approximation) the feedbacks of
gravity and electromagnetism. The corresponding equation of motion
has simply become the one of an spherically symmetric state on a
Minkowskian background, i.e.
\begin{equation}
\phi,_{rr} \, + \, \frac{2}{r}
\, \phi,_r \, - \,
\phi ,_{tt}
\, - \, m_0^2 \, \phi \, = \, 0  \; ,
\; \; {\rm and \; its \; h.c.} \; ,
\end{equation}
with the positive-frequency mode solutions
\begin{equation}
\phi \, = \, \frac{\cal N}{r} \,
e^{i(kr - \omega_k t)} \; \;   \Rightarrow \; \;
\bar{\phi} \, = \, \frac{\bar{\cal N}}{r} \,
e^{-i(kr - \omega_k t)} \, ,
\end{equation}
where $\omega_k \, = \, \left[ k^2 + m_0^2 \right]^{1/2}$.
Moreover, by neglecting the gravity feedback
on the Maxwell sector also,
as it effectively involves second-order contributions of
the charged scalar $\phi$,
the Lorentz condition and the Maxwell equations (5)
are satisfied, in the minimally
symmetric ansatz $A_1 = A_1(r,t)$, $A_4 = A_4(r,t)$,
$\phi = \phi (r,t)$, by the solution(s)
\begin{eqnarray}
A_1 & = & e k | {\cal N}|^2 \nonumber \\* A_4 & = & 2e \omega_k
|{\cal N}|^2 \, \log \frac{r}{r_0} \, + \, 2ek \, \frac{|{\cal
N}|^2}{r} \, t \; ,
\end{eqnarray}
which correspond to the electric field
\begin{equation}
E \, = \, F_{14} \, \approx \,
A_{4,r} - A_{1,t} \, = \,
2e \, \omega_k \,
\frac{|{\cal N}|^2}{r} - 2ek \,
\frac{|{\cal N}|^2}{r^2} \, t
\end{equation}

In the particular case $k=0$, where the star is just above the
passage to the possible stable excited states, its mode pulsation
$\omega = m_0$ being located at the accumulation point of the
eigenfrequencies of an excited boson star, we have found for the
linearized Einstein field equations
\begin{eqnarray}
& &
\frac{2}{r} h,_r \, - \, \frac{2}{r^2} f
= \kappa \left[
\frac{|{\cal N}|^2}{r^4} \, - \, 2e^2
m_0^2 \, \frac{|{\cal N}|^4}{r^2} \, + \,
4e^2 m_0^2 \, \frac{|{\cal N}|^4}{r^2} \,
\log \frac{r}{r_0} \right] ;
\nonumber \\*
& &
h,_{rr} \, + \, \frac{1}{r} \left( h,_r - f,_r \right) =
\kappa \left[
- \, \frac{|{\cal N}|^2}{r^4} \, + \, 2e^2
m_0^2 \, \frac{|{\cal N}|^4}{r^2} \, + \,
4e^2 m_0^2 \, \frac{|{\cal N}|^4}{r^2} \,
\log \frac{r}{r_0} \right] ;
\nonumber \\*
& &
\frac{2}{r} f,_r \, + \, \frac{2}{r^2} f
= \kappa \left[ 2 m_0^2 \,
\frac{|{\cal N}|^2}{r^2} \, + \,
\frac{|{\cal N}|^2}{r^4} \, + \,
2 \, e^2 m_0^2 \, \frac{|{\cal N}|^4}{r^2} +
4 e^2 m_0^2 \, \frac{|{\cal N}|^4}{r^2}
\log \frac{r}{r_0} \right] \nonumber \\*
\end{eqnarray}
the solutions
\begin{eqnarray}
f(r) & = & \frac{C_1}{r} \, - \, \frac{b}{2r^2} \, + \, a \, \log
\frac{r}{r_0}  \; , \nonumber \\* h(r) & = & - \; \frac{C_1}{r} \,
,
\end{eqnarray}
where
\begin{equation}
b \, = \, \kappa \, |{\cal N}|^2 \; , \; \; a \, = \, m_0^2 \, b
\; , \; \; {\rm and} \; \; |{\cal N}|^2 = \frac{1}{2 e^2} \, ,
\end{equation}
that have led to analytical expressions for total charge, particle
number, radius and mass of the analyzed configuration [7].

In the followings, we are going further and analyze the feedback
of gravity and electric field, respectively expressed by the
metric functions (13) and the four-potential (10), on the
Klein-Gordon equations (4). Within a first-order perturbative
approach, we write down the wave function describing the charged
scalar field as
\begin{equation}
\Phi (r, \theta , \varphi , t) \, = \,
\phi(r,t) + \chi (r, \theta , \varphi ,t) \, ,
\end{equation}
where $| \chi | \ll | \phi |$ and
$\phi$ is the Minkowskian background
solution (9), with $|{\cal N}|=1/ \sqrt{2e^2}$.
Consequently,
the Klein-Gordon equations (4), with
$k=0$ implying $\omega =m_0$, turn into
\begin{eqnarray}
& & e^{\frac{b}{r^2}-\frac{2 C_1}{r}}
\left[ \frac{\partial^2 \chi}{\partial r^2} +
\frac{2}{r} \, \frac{\partial \chi}{\partial r} \right]
+ \frac{1}{r^2} \, \tilde{\Delta} \chi - e^{\frac{2C_1}{r}}
\frac{\partial^2 \chi}{\partial t^2} - m_0^2 \left[
1- \log^2 \left( \frac{r}{r_0} \right) \right] \chi
\nonumber \\*
& & = \; - \, \frac{2C_1}{r^2} \, e^{\frac{b}{r^2}
- \frac{2C_1}{r}} \; \frac{\partial \chi}{\partial r}
- 2 i m_0 e^{\frac{C_1}{r}} \, \log \left(
\frac{r}{r_0} \right) \frac{\partial \chi}{\partial t}
\nonumber \\*
& & + \; e^{-i m_0 t} \left \lbrace \frac{m_0^2
\left( 1- e^{\frac{C_1}{r}} \right)}{\sqrt{2} er}
- \frac{\sqrt{2} m_0^2}{er} \left[ 1+ \log \sqrt{
\frac{r}{r_0}} \right] \log \left( \frac{r}{r_0} \right)
\right \rbrace
\end{eqnarray}
and its h.c., and we shall focus on the long range behaviour, $r_0
\ll r < \infty$, where the above equation gets the simpler form
\begin{eqnarray}
& & e^{-\frac{2 M}{r}} \left[ \frac{\partial^2 \chi}{\partial r^2}
+ \frac{2}{r} \, \frac{\partial \chi}{\partial r} \right] +
\frac{1}{r^2} \, \tilde{\Delta} \chi - e^{\frac{2M}{r}} \,
\frac{\partial^2 \chi}{\partial t^2} \nonumber \\* & & + \, 2 i
m_0 \, e^{\frac{M}{r}} \log \left( \frac{r}{r_0} \right)
\frac{\partial \chi}{\partial t} + m_0^2 \log^2 \left(
\frac{r}{r_0} \right) \chi \nonumber \\* & & = \, - \,
\frac{m_0^2}{\sqrt{2} \, er} \log^2 \left( \frac{r}{r_0} \right)
e^{-i m_0 t}
\end{eqnarray}
The constant $C_1$ has been replaced by the total gravitational
mass of the Bose star, given by the Tolman's relation [13]
\begin{equation}
M \, = \, \int T_{44} \, e^{f+h} \, 4 \pi r^2 dr \; ,
\end{equation}
and explicitly reading [7]
\begin{eqnarray}
M & = & \frac{2 \pi}{\kappa} \,
\sqrt{ 2b \left( \frac{b}{2} \right)^a} \left \lbrace
\Gamma \left( \frac{1-a}{2} \right) \right.
\nonumber \\*
& + &
\left. \frac{a}{2} \, \Gamma \left( - \, \frac{1+a}{2}
\right) \left[ 3 + \log \frac{b}{2} -
PolyGamma \left( 0, - \, \frac{1+a}{2}
\right) \right] \right \rbrace
\end{eqnarray}
Since, in the radiation zone, one can consider
\[
e^{\pm \frac{2M}{r}} \approx 1 \pm \frac{2M}{r} + {\cal O} \left[
\left( \frac{2M}{r} \right)^{n \geq 2} \right] ,
\]
the equation (17) does actually become
\begin{eqnarray}
\Box \chi & = &
\frac{2M}{r} \left \lbrace
\frac{1}{r^2} \frac{\partial \;}{\partial r}
\left[ r^2 \frac{\partial \chi }{\partial r} \right]
+ \frac{\partial^2 \chi}{\partial t^2} \right \rbrace
\nonumber  \\*
& - &
2i m_0 \log \left( \frac{r}{r_0} \right)
\frac{\partial \chi}{\partial t} \, -
\, \frac{m_0^2}{\sqrt{2} er} \, e^{-im_0 t} \log^2
\left( \frac{r}{r_0} \right) ,
\end{eqnarray}
where $\Box$ is the usual d'Alembertian on ${\bf{\rm R}}^4$, and
therefore, once it achieved the standard form,
\begin{equation}
\Box \chi \, = \, \hat{V} \chi + {\cal J} \, ,
\end{equation}
employed in Perturbation Theory, it points out the operators
\begin{eqnarray}
\hat{V} (r,t) & = &
\frac{2M}{r} \left \lbrace
\frac{1}{r^2} \frac{\partial \;}{\partial r}
\left[ r^2 \frac{\partial \;}{\partial r} \right]
+ \frac{\partial^2 \;}{\partial t^2} \right \rbrace
- 2i m_0 \log \left( \frac{r}{r_0} \right)
\frac{\partial \;}{\partial t} \; ,
\\*
{\cal J} (r,t) & = &
- \, \frac{m_0^2}{\sqrt{2} er} \,
e^{-im_0 t} \log^2 \left( \frac{r}{r_0} \right)
\end{eqnarray}
describing the (perturbed) effective potential and current. The
{\it massless-like} 0-th order equation
\begin{equation}
\frac{1}{r^2} \frac{\partial \;}{\partial r} \left( r^2 \,
\frac{\partial h}{\partial r} \right) + \frac{1}{r^2} \,
\tilde{\Delta} h - \frac{\partial^2 h}{\partial t^2} \, = \, 0
\end{equation}
provides the complete orthonormal set of positive-frequency modes
(in terms of spherical Hankel functions)
\begin{equation}
h_{\omega lm} (r, \theta , \varphi , t) \, = \, \frac{1}{2
\sqrt{r}} \, H_{l+\frac{1}{2}}^{(1)} ( \omega r) Y_l^m (\theta ,
\varphi) e^{-i \omega t} \, ,
\end{equation}
which enables us to compute the first-order transition amplitudes
between the initial and final states as:
\begin{equation}
{\cal A}_{\omega lm}^{\omega^{\prime} l^{\prime} m^{\prime}}
= \int h^*_{\omega^{\prime} l^{\prime} m^{\prime}} (x)
 \left( \hat{V} h_{\omega lm}
(x) \right) r^2 dr d \Omega dt \, ,
\end{equation}
where
\begin{equation}
\hat{V} h_{\omega lm} (x) \, = \, - \, 2 \left \lbrace \frac{M}{r}
\left[ 2 \omega^2 - \frac{l(l+1)}{r^2} \right] + \, 2 m_0 \,
\omega \log \left( \frac{r}{r_0} \right) \right \rbrace h_{\omega
lm} (x) \; ,
\end{equation}
with $(x) = (r, \theta , \varphi ,t )$. In (26), we identify the
transitions of the charged boson in the presence of gravity
\begin{eqnarray}
{\cal A}_{SO(3,1)}^I & = &
-  \, M \omega^2 \, \int_{r \to 0}^{r \to \infty}
H_{l+\frac{1}{2}}^{(2)} (\omega^{\prime} r) \,
H_{l+ \frac{1}{2} }^{(1)} (\omega r) \, dr
\nonumber \\*
& = &
\frac{2 \gamma}{\pi} \, M \omega \; ,
\; \; {\rm for } \; \; l=0 , \, 1 \, ;
\nonumber \\*
{\cal A}_{SO(3,1)}^{II} & = &
\frac{M l(l+1)}{2} \,
\int_{r \to 0}^{r \to \infty} \frac{dr}{r^2} \,
H_{l+\frac{1}{2}}^{(2)} (\omega^{\prime} r)
H_{l+\frac{1}{2}}^{(1)} (\omega r)
\nonumber \\*
& = &
\frac{M \omega}{\pi} \, ,
\end{eqnarray}
and of the electric field potential
\begin{eqnarray}
{\cal A}_{U(1)} & = &
- \, m_0 \,
\omega \int_{r \to 0}^{r \to \infty}
H_{l+\frac{1}{2}}^{(2)} (\omega^{\prime} r)
\left[ r \log \left( \frac{r}{r_0} \right) \right]
H_{l+\frac{1}{2}}^{(1)} (\omega r) \, dr
\nonumber \\*
& = &
- \, \left[ \frac{m_0}{\omega} \left(
l + \frac{1}{2} \right) \right] ,
\end{eqnarray}
where
$\gamma$ is the Euler's constant $\gamma
\approx 0.577216$ and the well-known $\delta$-Dirac
multiplier $2 \pi \delta ( \omega^{\prime} - \omega )$
has been factored out.

The corresponding transition rate for each
of the above amplitudes is respectively given by
\begin{eqnarray}
& &
\frac{d \;}{dt} {\cal P}_{SO(3,1)}^I
\, = \,
2 \pi \delta ( \omega^{\prime} - \omega ) \,
\left( \frac{2 \gamma}{\pi} \right)^2
( M \omega^2 )^2
\nonumber \\*
& &
\frac{d \;}{dt} {\cal P}_{SO(3,1)}^{II}
\, = \,
2 \pi \delta ( \omega^{\prime} - \omega ) \,
\left( \frac{M \omega^2}{\pi} \right)^2
\nonumber \\*
& &
\frac{d \;}{dt} {\cal P}_{U(1)}
\, = \,
2 \pi \delta ( \omega^{\prime} - \omega ) \,
m_0^2 \left( \frac{\omega^{\prime}}{\omega} \right)^2
\left( l + \frac{1}{2} \right)^2
\end{eqnarray}

Finally, let us turn to the current operator (23) in order to
study the possibility of spontaneous creation of charged bosons in
the presence of the electric field potential $A_4$. These process
is described by the transition amplitude
\begin{eqnarray}
{\cal A}_{\cal J} & = &
\int \sqrt{\omega} \, h^*_{\omega lm} (x)
{\cal J}(r,t) \, r^2 dr d \Omega dt
\nonumber \\*
& = & - \, 2 \pi \delta (\omega -m_0 ) \,
\sqrt{2 \pi \omega} \,
\frac{m_0^2}{e} \,
\int_{r \to 0}^{ r \to \infty}
\sqrt{r} \, H_{1/2}^{(2)} (\omega r) \log^2
\left( \frac{r}{r_0} \right)
\end{eqnarray}
Integrating by parts with a cut-off in
$z \equiv r/r_0$, the previous expression becomes
\begin{equation}
{\cal A}_{\cal J} \, = \, 2 \pi \delta (\omega -m_0 ) \,
\frac{2m_0^2}{e \omega} \, {\cal I} \; , 
\end{equation}
where the integral
\begin{eqnarray}
{\cal I} & = & \int_1^{\infty} \frac{dz}{z} \, e^{-ibz} \, \log
(z) \nonumber \\* & = & - \, \frac{\pi^2}{24} + \frac{\gamma^2}{2}
+ \frac{i}{2} \gamma \pi + \left[ \gamma + \frac{i \pi}{2} +
\frac{\log (b)}{2} \right] \log (b) \nonumber \\* & & - \,
\frac{b^2}{8} \,_3 F_4 \left[ \lbrace 1,1,1 \rbrace , \left
\lbrace \frac{3}{2} , 2, 2, 2 \right \rbrace , \, - \,
\frac{b^2}{4} \right] \nonumber \\* & & - \, i \, b \,_2 F_3
\left[ \left \lbrace \frac{1}{2}, \frac{1}{2} \right \rbrace ,
\left \lbrace \frac{3}{2} , \frac{3}{2}, \frac{3}{2} \right
\rbrace , \, - \, \frac{b^2}{4} \right] ,
\end{eqnarray}
with $b \equiv \omega r_0$, contains the
generalized hypergeometric functions \\
$\,_p F_q \left[ \left \lbrace a_1 , \dots , a_p
\right \rbrace , \left \lbrace b_1 , \dots , b_q
\right \rbrace , z \right]$ which are
finite for all finite arguments and $p \leq q$
[14].
In the above calculations we have assumed that
$\omega = 2 \pi/T$ is the natural dual
of the ^^ ^^ repeating time-interval'' such that
\[
\lim_{T \to \infty} \int_0^T \int_{{\rm{\bf R}}_+}
\frac{dt \, d \omega}{2 \pi} \, = \,
\Sigma_{states} 1 \, ,
\]
and then,
\begin{equation}
\frac{d \;}{dt} {\cal P}_+ (m_0 ; {\cal J} )
\, = \,
 2 \pi \delta (\omega -m_0 ) \,
\left( \frac{2m_0^2}{e \omega} \right)^2
\left| {\cal I} \right|^2
\end{equation}
does concretely represent the coherent
source-field regeneration rate.

The authors wish to acknowledge the kind hospitality and fertile
environment of the University of Oregon where this work has been
carried out. M.A.D. thanks the U.S. Department of State, the
Council for International Exchange of Scholars (C.I.E.S.) and the
Romanian-U.S. Fulbright Commission for sponsoring her
participation in the Exchange Visitor Program no. G-1-0005.
Documentation on boson stars provided by Ph. Jetzer, Scott Hawley,
Ed Seidel and Marcelo Gleiser has been of a real help. Professors'
N.G. Deshpande and S. Hsu inciting discussions are highly
regarded.

\end{document}